\documentclass[prl,aps,twocolumn,floatfix,superscriptaddress]{revtex4-2}
\usepackage{amsfonts,amsmath,amsthm,amssymb,mathtools,bm}
\usepackage{graphics,color,float,multirow}
\usepackage{microtype}
\usepackage{hyperref}

\begin{document}

\title{Elucidating the mechanism of helium evaporation from liquid water}
\author{Kritanjan Polley}
\affiliation{Chemical Sciences Division, Lawrence Berkeley National Laboratory, Berkeley, CA 94720, USA}
\affiliation{Department of Chemistry, University of California, Berkeley, CA 94720, USA}
\author{Kevin R. Wilson}
\affiliation{Chemical Sciences Division, Lawrence Berkeley National Laboratory, Berkeley, CA 94720, USA}
\author{David T. \surname{Limmer}}
\email{dlimmer@berkeley.edu}
\affiliation{Chemical Sciences Division, Lawrence Berkeley National Laboratory, Berkeley, CA 94720, USA}
\affiliation{Department of Chemistry, University of California, Berkeley, CA 94720, USA}
\affiliation{Materials Sciences Division, Lawrence Berkeley National Laboratory, Berkeley, California 94720, USA}
\affiliation{Kavli Energy NanoScience Institute, Berkeley, California 94720, USA}

\begin{abstract}
We investigate the evaporation of trace amounts of helium solvated in liquid water using molecular dynamics simulations and theory. Consistent with experimental observations, we find a super-Maxwellian distribution of kinetic energies of evaporated helium. This excess of kinetic energy over typical thermal expectations is explained by an effective continuum theory of evaporation based on a Fokker-Planck equation, parameterized molecularly by a potential of mean force and position-dependent friction. Using this description, we find that helium evaporation is strongly influenced by the friction near the interface, which is anomalously small near the Gibbs dividing surface due to the ability of the liquid-vapor interface to deform around the gas particle. Our reduced description provides a mechanistic interpretation of trace gas evaporation as the motion of an underdamped particle in a potential subject to a viscous environment that varies rapidly across the air-water interface. From it we predict the temperature dependence of the excess kinetic energy of evaporation, which is yet to be measured.
\end{abstract}

\maketitle

Air-water interfaces mediate a number of important physical and chemical processes. One process of primary importance to atmospheric chemistry is the uptake of soluble gases.\cite{limmer24,wilson24} Gas molecules in the air have to pass through the air-water  interface to become solvated into the bulk liquid, or equivalently, volatile species solvated in solution must cross through the interface to evaporate. For inert species, the mass transfer between the solution and the air is a physical process that requires diffusion across an environment that varies rapidly on molecular lengthscales.\cite{singh2022peptide,shin2018three} Recently, we have developed a theoretical framework to systematically coarse-grain the molecular dynamics accompanying gas uptake using effective continuum equations, parameterized with molecularly resolved forces that serve to dissipate energy and those that establish steady-state concentration profiles.\cite{polley24} Using this framework we revisit the evaporation of trace helium from liquid water, providing a mechanism for the emergence of a super-Maxwellian distribution of kinetic energy. 

It has been observed both experimentally and with molecular simulation that helium exhibits a non-Maxwell velocity distribution as it evaporates from the air-water interface.\cite{nathanson2021liquid} This phenomenon has been observed consistently in experiments with pure water, salt-water solutions, and in the presence of surfactants.~\cite{hahn16,kann16,gao24} This behavior is  anomalous as the evaporation of water as well as most solutes display a Maxwell-Boltzmann distribution, consistent with thermal equilibrium.\cite{varilly13} Aqueous helium solutions however are not the only systems to exhibit this effect, helium also shows this behavior from hydrocarbon liquids like squalane, dodecane, octane, isooctane, ethylene glycol solutions, and jet fuels.~\cite{johnson14,williams15,lancaster15} Moreover, non-Maxwell energy distributions have been observed for carboxylic acid dimer evaporation from water surface.~\cite{faubel89} 

Most theoretical studies on helium evaporation have rationalized the super-Maxwellian velocity distribution using the potential of mean force of helium near the interface.~\cite{hahn16,kann16,phillips08}  Helium has a large, positive solvation free energy, and corresponding small Henry's law constant, due to its weak dispersion interactions with the solvent. Further, the potential of mean force decreases sharply from the liquid, as helium exits into the vapor. 
The subsequent force derivable from the potential of mean force is envisioned as pushing a helium atom out of solution. Indeed,
the temperature and composition dependence of the excess kinetic energy accompanying helium evaporation over thermal expectations have been correlated with the changes to the solvation free energy, with larger more endothermic solvation resulting in larger kinetic energies accompanying evaporation \cite{hahn16}. However, this description is necessarily incomplete, as the potential of mean force encodes only the average force on a molecule moving infinitesimally slowly \cite{limmer2024statistical}. At any finite velocity, dissipative, frictional forces in addition to conservative forces from the potential of mean force are needed to accurately describe particle motion. 

Here, we investigate the nature of helium evaporation by including friction and its variation near the interface into the description of its dynamics. This is made possible by recent algorithmic advances for computing the spatially dependent friction opposing motion perpendicular to an extended interface \cite{polley24}. Like the potential of mean force, the friction varies rapidly between the liquid and vapor sides of the interface, and by including it into a Fokker-Planck equation \cite{zwanzig2001nonequilibrium}, we are able to reproduce both experimental observations as well as detailed calculations from simulation.

\begin{figure*}[t]
    \centering
    \includegraphics[width=17cm]{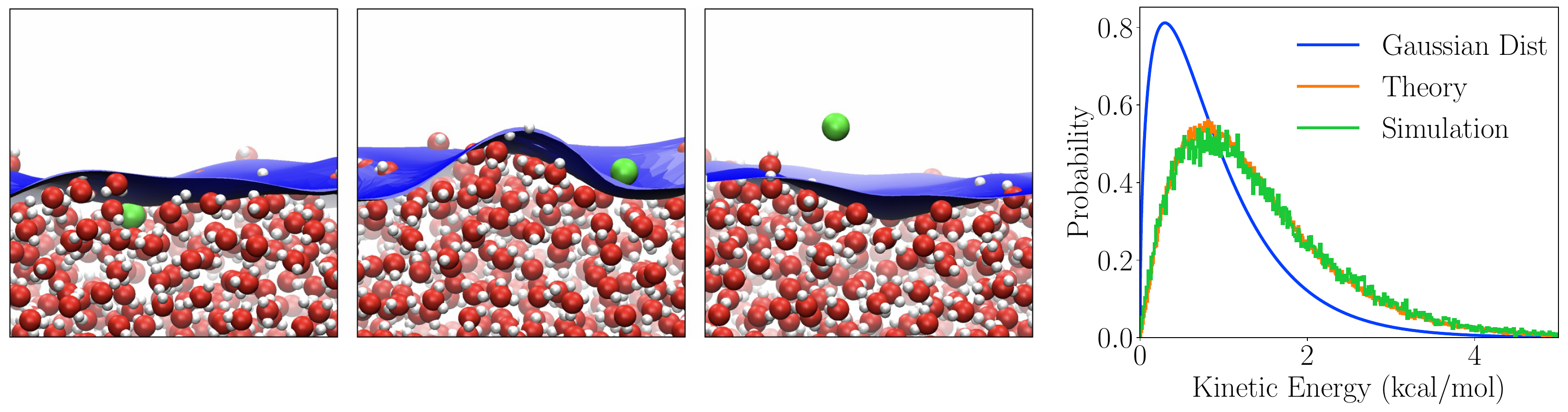}
    \caption{Snapshots of He (green) fixed near the air-water interface ($z=16, 18, \,\text{and} \, 22 \, \text{\r{A}}$, respectively from left to right). The instantaneous interface is shown in blue. After $z=18 \text{\r{A}}$, helium mostly stays above the instantaneous interface. The kinetic energy distribution of the evaporated helium is shown on the right in orange. A distribution for a particle with a Gaussian distribution is shown in blue. The kinetic energy distribution from molecular dynamics trajectories is displayed in green. Results in this figure are at 300 K.}
    \label{figInstInterface}
\end{figure*}

To study helium evaporation, molecular dynamics simulations were first performed. Molecular dynamics simulations were carried out in a slab geometry with a polarizable force fields in LAMMPS~\cite{thompson22}. A slab containing 768 water molecules and one helium atom at 300 K was used, embedded in a domain with dimensions $24.8 \times 24.8 \times 111.8 \,\mathrm{\AA}^3$ where the larger dimension, aligned along the $z$-axis, is perpendicular to the interface. Periodic boundary conditions were applied in all directions. The SWM4-NDP~\cite{lamoureux06} water model was used in this study, which combines polarizability with Lennard-Jones non-bonded interactions and rigid-body dynamics.~\cite{miller02} The helium-water interactions were described similarly with polarizability for helium equal to 0.205 $\mathrm{\AA}^3$ and Lennard-Jones parameters derived from Lorentz-Berthelot mixing rules with helium parameters take from Ref.~\onlinecite{halgren92}. Polarizability was introduced using a Drude oscillator model~\cite{lemkul16,huang14,savelyev14} along with a symmetrization procedure described by Dodin and Geissler.~\cite{dodin23} A particle-particle-particle-mesh method~\cite{pollock96} was used for the long range Coulomb interactions with a target relative error of 10$^{-5}$. The Lennard-Jones interactions were truncated and shifted at a distance of 12 $\mathrm{\AA}$. 

The simulations we perform initialize helium in the bulk of the liquid and propagate dynamics until the helium atom evaporates. Representative snapshots of this process are shown in Fig.~\ref{figInstInterface}, where the interface is oriented perpendicular to the $z$-axis. The initial position distribution for the helium atom is sampled uniformly, within a 10 \AA\ region from the interface in the bulk phase, well below the local instantaneous interface,\cite{willard10} and the helium is considered evaporated when it is 20 \AA\ away from the local instantaneous interface in the direction of the vapor.\cite{varilly13} 
The kinetic energy distribution of helium upon evaporation are displayed in the rightmost panel of Fig.~\ref{figInstInterface}. Consistent with previous reports, we reproduce the non-Maxwellian  distribution.\cite{hahn16,kann16} We find a mean kinetic energy at 300 K of $1.50$ kcal/mol, an excess of $0.3$ kcal/mol over $2 k_\mathrm{B}T$ expected from the flux weighted Maxwell-Boltzmann distribution.\cite{kann16} The velocity in the $z$ direction is markedly non-Gaussian, while the velocity distribution along the direction parallel to the interface direction show slight deviations from a Maxwell-Boltzmann distribution. 

In order to rationalize this observation of a non-Maxwellian distribution, we employ our previously described procedure of coarse-graining the dynamics of the helium into a continuum Fokker-Planck equation.~\onlinecite{polley24} Within this reduced description the solvent and orthogonal degrees of freedom to the $z$-axis are integrated out, resulting in a stochastic differential equation with which we can study the evaporation process. This statistical description incorporates molecular details from atomistic simulation in the form of friction and free energy, into macroscopic mass transport equations. The resultant equation is 
\begin{align}\label{eqFPE}
    \frac{\partial p}{\partial t}=& -v_z\frac{\partial p }{\partial z}+ \frac{\partial}{\partial v_z} \left [ \frac{\partial_z F(z)}{m}+ \gamma(z) v_z \right ]p \\
    &+ \frac{k_{\mathrm{B}}T\gamma(z)}{m} \frac{\partial^2  p}{\partial v_z^2}\nonumber
\end{align}
where $p(z,v_z,t)$ is the probability of finding the helium at position $z$ with velocity $v_z$ at time $t$. $F(z)$ is the potential of mean force, $m$ is the mass of helium, and $\gamma(z)$ is the mass-weighted friction. The system is described at constant temperature $T$, and $k_{\mathrm{B}}$ is Boltzmann's constant. We use a time-local equation, thereby assuming that the equation of motion is Markovian and the solvent degrees of freedom relax quickly. 

\begin{figure*}[t]
    \includegraphics[width=17cm]{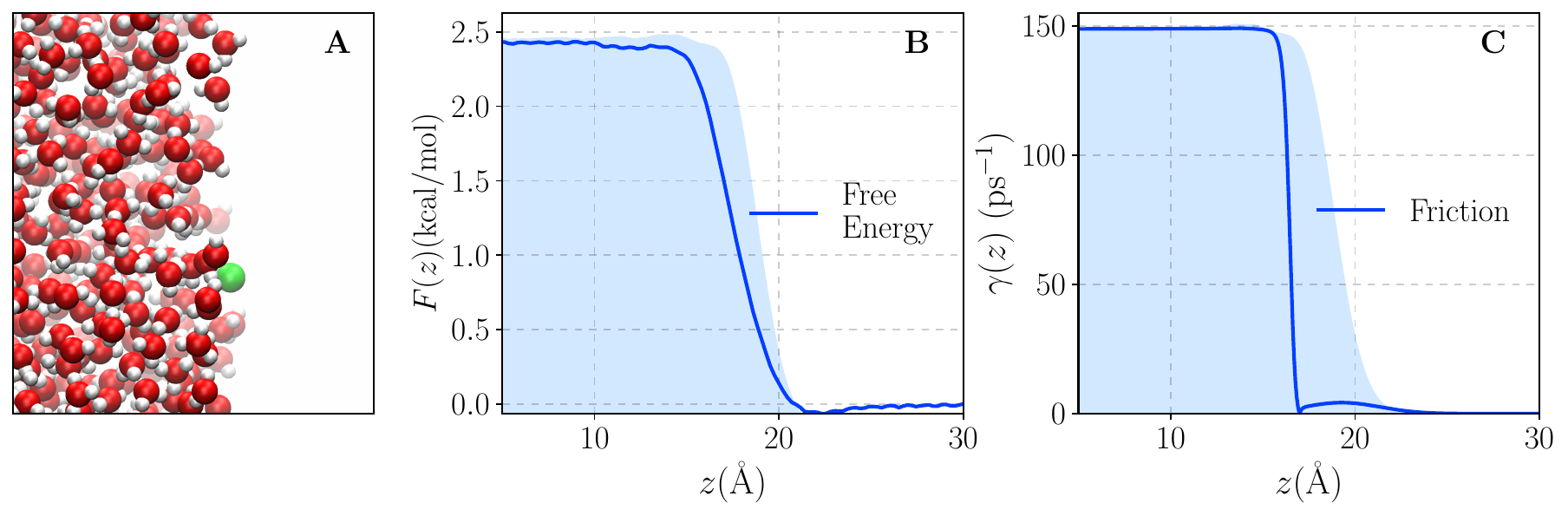}
    \caption{(A) A snapshot of MD simulation of He near the interface. We have adopted the same color scheme for molecules as in Fig.~\ref{figInstInterface}. (B) The free energy profile (dark blue) for moving He through the air-water interface. (C) The variation in position dependent friction (dark blue) of He for this system. The shaded light blue region shows the water density profile (scaled) in both panels B and C.}
    \label{fighePMF}
\end{figure*}
To parameterize Eq.~\ref{eqFPE} we require the potential of mean force, $F(z)$, and the spatially dependent friction, $\gamma(z)$. The potential of mean force, $F(z)$, is the reversible work to move the particle along the $z$ direction. We have computed it with umbrella sampling,~\cite{kastner11} where it is estimated by the probability of finding the helium atom with position $z_0$ at position $z$,
\begin{equation}
    F(z)=-k_{\mathrm{B}}T \ln \langle \delta(z-z_0)\rangle+F_0,\label{eqPMF}
\end{equation} 
where the angular bracket indicates a canonical average. The added constant, $F_0$, is a reference free energy that sets the free energy in the vapor phase to 0.  A harmonic bias potential of the form $k_z(z_0-\bar{z})^2$ is applied, where spring constant $k_z=2\,$ kcal/mol/\AA$^2$, centered at $\bar{z}$. A total of 71 locations for $\bar{z}$ from $-35$ \AA\, to $35$ \AA\, with a spacing of 1 \AA\, is employed within the aforementioned slab geometry. To sample the canonical distribution, an extended Lagrangian approach is used, incorporating a velocity-Verlet~\cite{frenkel23} time integration scheme. A small mass and kinetic energy are attributed to the Drude particles. The amplitude of the Drude oscillator is regulated using a low temperature thermostat at 1 K acting in the center-of-mass of each atom-Drude pair.~\cite{lamoureux03} A time step of 1 fs is used. Umbrella sampling data are collected from 15 ns production runs after equilibration for 1 ns. Different simulations are combined with the weighted histogram analysis method.~\cite{kumar92,whamCode} 

The potential of mean force for helium is depicted in Fig.~\ref{fighePMF}(B). The potential of mean force is nearly monotonically decreasing as a function of increasing $z$, with a nearly sigmoidal profile. Deviations from both features are due to the very slight affinity for the air-water interface that results in a very shallow minima just above the Gibbs dividing surface. The free energy of solvation, $F_{\mathrm{sol}}$, defined as the free energy difference between the plateau in the bulk liquid phase and the plateau in the bulk vapor phase, is $\sim 2.45$ kcal/mol which is in good agreement with experimental value of 2.78 kcal/mol.~\cite{clever15}

The spatially dependent friction is determined from the procedure described in Ref.~\onlinecite{polley24}. Briefly, an ensemble of short time molecular dynamics trajectories are integrated to produce an estimate of $p(z,v_z,t)$, which is then compared to the numerical solution of Eq. \ref{eqFPE} with a trial function for $\gamma(z)$. The Fokker-Planck equation is numerically solved by 
integrating on a two-dimensional grid with a fourth-order Runga-Kutta integrator.
We start with a functional form for the friction with appropriate asymptotes in the gas and liquid phases. The liquid and gas phase frictions for helium are obtained using the Einstein relation, $\gamma_{l,g}=k_{\mathrm{B}}T/mD_{l,g}$, with $D_l=4.19\times10^{-5}$ cm$^2$/s and $D_g=0.697$ cm$^2$/s.~\cite{fuller66} The bulk phase diffusion constants for this system are obtained from the velocity auto-correlation function~\cite{frenkel23} and are close to experimentally observed values.~\cite{verhallen84} The function that interpolates between these asympotes is described by a set of adjustable parameters. In our previous work, we used a scaled sigmoid function, with shape parameters describing its center, width and compression. In the present study of helium evaporation, the sigmoid form was not adequate and we introduced additional Gaussian functions around the inflection point in the sigmoid curve to allow further variation in its form.  The parameters of the interpolation function are then optimized using a simulated annealing procedure~\cite{goffe94}, as implemented in Julia,~\cite{mogensen18} to match the  distribution from the molecular dynamics simulation with the distributions generated from Eq.~\ref{eqFPE}. The initial condition for helium is a narrow Gaussian distribution in position, at $z=14$ \AA\, (\textit{c.f.} Fig.~\ref{fighePMF}), and a Maxwell-Boltzmann distribution in velocity. We ran 40,000 explicit MD trajectories for 30 ps to generate the position and velocity distributions. Data are collected at an interval of 1 ps between 8 and 30 ps and 0.1 ps between 0 and 8 ps. 

The optimized friction profile is displayed in Fig.~\ref{fighePMF}(C). Notably, the friction drops precipitously before the water density profile starts decreasing. This drop in friction is understood by the examination of instantaneous interfaces as depicted in Fig.~\ref{figInstInterface}.~\cite{willard10} As illustrated in the snapshots, helium spends time mostly above the instantaneous interface in the regions where the friction starts decreasing. So while helium is below the Gibbs dividing surface, it is thermodynamically favorable to deform the interface around it due to the large unfavorable solvation free energy. The free energy of helium starts also  changing before the water density profile decreases, but more gradually. The weak interaction between helium and water molecules compared to water-water interaction forces the helium to occupy the valleys in the air-water surface which keeps the overall water density relatively large, but the helium atom experiences a more gas-phase-like environment. Additional analysis shows that it is repulsive interactions that are overwhelmingly responsible for the form of the friction. Attractive interactions are negligible.  

Using $F(z)$ and $\gamma(z)$ computed from the molecular simulations, we have used the minimal model in Eq.~\ref{eqFPE} to calculate the distribution of kinetic energies for an evaporating helium atom. 
To compute the distribution of kinetic energies we integrate the associated Langevin equation for Eq.~\ref{eqFPE} with a coordinate dependent friction using the algorithm described in Refs.~\onlinecite{farago14a,farago14b,regev16}.
As with the molecular dynamics simulations, an initial distribution for the helium atom is sampled uniformly, within a 10 \AA\, region from the interface in the bulk phase. As we do not describe motion in the directions parallel to the interface within the continuum model, we assume that they stay distributed according to a Maxwell-Boltzmann distribution at 300 K. 

Shown in Fig.~\ref{figInstInterface}, the continuum model is able to reproduce the observed non-Maxwell distribution from the molecular dynamics simulations quantitatively. Both free energy and friction play a crucial roles in this result. The interfacial region in the potential of mean force does not provide any stability for helium and the friction decreases to almost gas phase value near the Gibbs dividing surface. The potential thus acts to repel helium from the liquid, and it faces negligible frictional forces that would otherwise serve to relax its velocity back to a Maxwell-Boltzmann distribution. The fact that the friction is nearly equal to its gas phase value at the location of the repulsive force means that this relaxation to equilibrium is minimized over the time helium spends at the interface. Despite the purely diffusive dynamics that serve to evaporate helium from water, using the continuum model we can thus deduce mechanistic insight into the process.

\begin{figure}
    \centering
    \includegraphics[width=8.5cm]{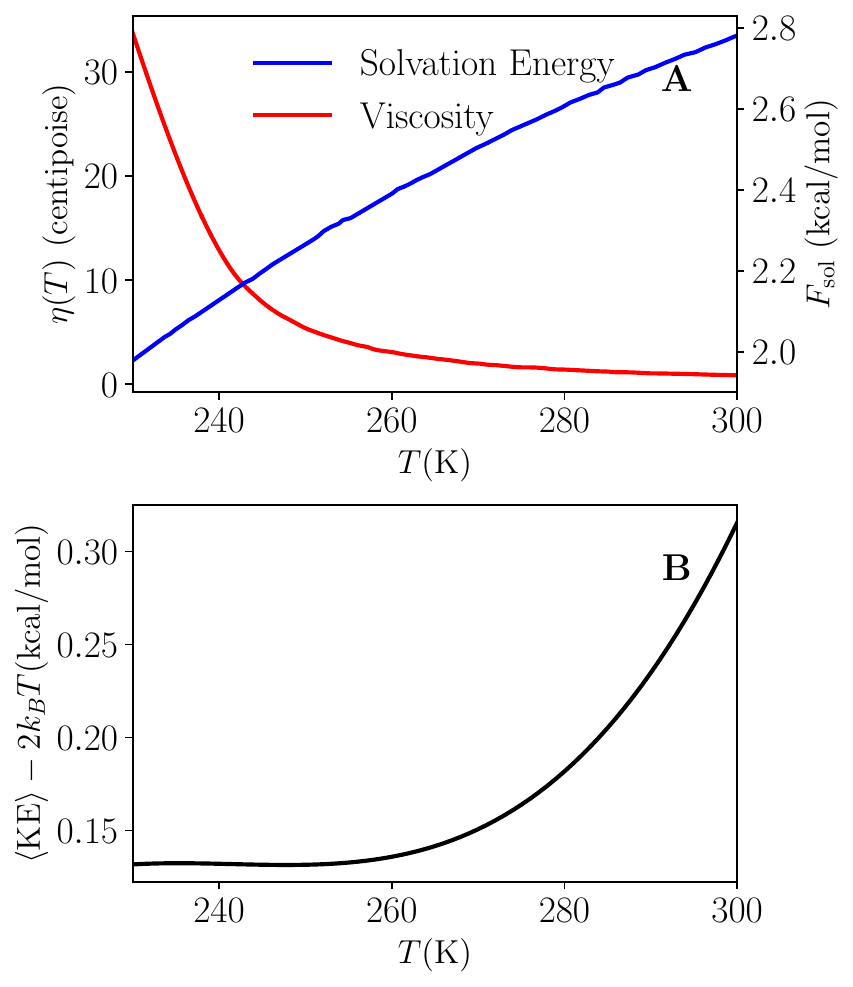}
    \caption{(A) Variation in viscosity of water (red) and helium solvation energy in water (blue) with temperature. Viscosity of supercooled water is taken from Ref.~\onlinecite{dehaoui15}, extrapolated below 239 K. The solvation energy data is taken from Refs.~\onlinecite{clever15,hahn16}. (B) A cubic fit of the excess mean kinetic energy of helium at different temperatures, collected 20 \AA\, away from the minima of the free energy profile in the vapor phase.}
    \label{figTempDepend}
\end{figure}

The quantitative agreement of our continuum model with the molecular dynamics simulation motivates its use to explore conditions in which molecular simulation is difficult. In particular, we have used the theory to make a prediction of the temperature dependence of the excess kinetic energy following evaporation. This is done by assuming that the spatial dependence of $F(z)$ and $\gamma(z)$ are temperature independent, but their amplitudes change with temperature. Experimentally, the temperature dependent solvation free energy $F_\mathrm{sol}(T)$ is known allowing us to approximate the free energy at $T$,  as $F(z,T)=F(z) \left [ F_\mathrm{sol}(T) -\Delta F \right ]/F_\mathrm{sol}(300 \mathrm{K})$ where $\Delta F$ is the difference between the simulated and experimental solvation energy at 300 K. Analogously the liquid phase friction is expected to follow a Stokes-Einstein relation, and since the temperature dependence of the viscosity of water, $\eta(T)$, is known experimentally, we can approximate the temperature dependent friction,  as $\gamma(z,T)=\gamma(z) \eta(T)/\eta(300 \mathrm{K})$. Both $F_\mathrm{sol}(T)$ and  $\eta(T)$ are shown in Fig.~\ref{figTempDepend}(A).

Following the same procedure detailed previously to compute the mean excess kinetic energy of evaporated helium, we can recompute its value for liquids that are kept at temperatures ranging from 230 K to 300 K. These are plotted in Fig.~\ref{figTempDepend}(B). The excess energy reaches a plateau at lower temperature where the friction increases sharply, and increases following the increasing solvation energy across the range of temperatures where the viscosity is roughly temperature independent. The temperature insensitively around 250 K is consistent with previous simulation results\cite{hahn16}, but this dependence stands as a testable prediction to be experimentally verified or refuted.

In this letter, we have investigated the abnormal evaporation of helium from an air-water interface with microscopic details coming from molecular simulation. The friction profile for helium decreases steeply near the interface, before the water density profile reaches its vapor phase value. The instantaneous interfaces for this system support the nature of the friction profile. The evaporation profile of helium computed from stochastic differential equations is in good agreement with results from explicit molecular dynamics simulation.  Application of the current method to systems of other solvents could be easily considered. For a small molecule like helium, one might expect nuclear quantum effects to be substantial,~\cite{hodges02} and a classical force field with polarization is only an approximation. Further investigation of the role of the solvent molecules near the interface, especially their fluctuation, will provide us important details in evaporation processes. 

\textbf{Data Availability : } Codes for generating the figures in this letter and the corresponding data are available at \href{https://zenodo.org/doi/10.5281/zenodo.13334084}{https://zenodo.org/doi/10.5281/zenodo.13334084}

\textbf{Acknowledgement : } The authors would like to thank Gil Nathanson for discussions on the early part of this work. This work is supported by the Condensed Phase and Interfacial Molecular Science Program (CPIMS) of the U.S. Department of Energy under Contract No. DE-AC02-05CH11231.

\bibliography{reference}

\end{document}